\documentclass[amsfonts,aps,pre,twocolumn,array,amsmath]{revtex4}

\usepackage{graphicx}

\newcommand{\be} {\begin{equation}}
\newcommand{\ee} {\end{equation}}
\newcommand{\bea} {\begin{eqnarray}}
\newcommand{\eea} {\end{eqnarray}}

\begin{document}
\title{Trap Models and Slow Dynamics in Supercooled Liquids}
\author{R. Aldrin Denny,$^1$ David R. Reichman,$^1$ \footnote[2]{Email: reichman@fas.harvard.edu} and  Jean-Philippe Bouchaud$^2$}
\affiliation{$^1$ Harvard University, 12 Oxford Street, Cambridge, Massachusetts 02138.}
\affiliation{$^2$ Service de Physique de l'\'Etat Condens\'e, Center d'\'etudes de
Saclay, Orme des Merisiers, 91191 Gif-sur-Yvette C\'edex, France.}
\date{\today}
\begin{abstract}
The predictions of a class of phenomenological trap models of
supercooled liquids are tested via computer simulation of a model
glass-forming liquid.  It is found that a model with a Gaussian
distribution of trap energies provides a good description of the
landscape dynamics, even at temperatures above $T_{c}$, the critical
temperature of mode-coupling theory.  A scenario is discussed whereby
deep traps are composed of collections of inherent structures above
$T_{c}$ and single inherent structures below $T_{c}$.  Deviations from
the simple Gaussian trap picture are quantified and discussed.\\
\end{abstract}
\maketitle

Glass-forming systems are ubiquitous in nature, and constitute the
basis for wide-ranging technological applications \cite{sci95,ean96}.
Furthermore, the theoretical and computational techniques developed
for the study of glass-forming liquids have been extremely useful in
the study of static and dynamic phenomena in other complex systems, such
as spin-glasses, polymers and proteins \cite{spin98}.  Unfortunately,
despite recent progress, the underlying microscopic origin of slow
dynamics in supercooled liquids remains largely an unsolved problem.

The mode-coupling theory ({\sc mct}), pioneered by G\"{o}tze and co-workers,
is unique in that quantitative predictions are made for the viscous
slowing-down exhibited by dynamical correlators with the input of
structural information alone \cite{gs92}.  In this sense, {\sc mct}
is perhaps the only {\em ab initio} theory of slow dynamics in supercooled
liquids.  {\sc mct} appears to be a quantitative theory of cage formation
at intermediate times, but has some shortcomings in describing cage
relaxation at long times.  In particular, {\sc mct} overestimates the
the cage effect, leading to a ``glass-transition'' at a temperature $T_{c}
> T_{g}$ where $T_{g}$ is the experimentally determined glass transition
temperature. Theoretical attempts have been made to include ``hopping''
processes that restore ergodicity below $T_{c}$, however such theories
currently demand a level of approximation that reduces the ability to
make quantitative predictions \cite{gs92}.

A seemingly different perspective is provided by the ``landscape'' picture of
slow dynamics in supercooled liquids \cite{s95,g69}.  Here, one attributes
slow structural relaxation to the complex pathways that connect
configurational states or ``inherent structures''({\sc is}) on the
multidimensional potential surface.  Wales and coworkers have
performed extensive studies of the details of the landscape properties
of model glass forming liquids \cite{wdmmw00,mw01}.  In particular, they
have noted that the landscape of Lennard-Jones liquids has a multi-funnel
structure. The inherent structures residing inside a funnel are separated
by small barriers, while barriers separating funnels may be large.  Heuer
and coworkers \cite{bh00,bh99,dhcm} have recently shown that the phase point
of an {\em unbiased} trajectory is trapped in a ``meta-basin'' for long
periods of time, making frequent small hops within the meta-basin, and
infrequent excursions from one meta-basin to another, a scenario that
that resonates with the multi-funnel picture.

On the other hand, recent studies of mean field p-spin glasses \cite{spin98}
have established a deep connection between the {\sc mct} and landscape
pictures. The {\sc mct} transition temperature $T_c$ appears as the point
below which the only available saddle points of the energy are minima,
whereas above $T_c$, the vast majority of saddles are higher order saddles
\cite{kl96}. Within {\sc mct} (or mean field p-spin glasses), the dynamics
freezes at $T_c$ in the lowest available `critical' saddle that percolates
in phase-space, but never penetrates the lower minima region below $T_c$.
This is related to the above mentioned inability of {\sc mct} to describe
hopping events.

The relevance of this result for Lennard-Jones systems was discussed in the
important work of Angelani {\em et al.} \cite{adrss00} and Broderix, Cavagna
{\em et al.} \cite{bbczg00,c01}, where the critical {\sc mct} temperature
$T_c$, as identified from the extrapolated divergence of dynamical time
scales, coincides with the point where the number of unstable saddles
vanishes. However, due to the non mean-field nature of the system, several
important differences with {\sc mct} appear. First, because barriers are
finite, the dynamics below $T_c$ is now dominated by activated transitions
between basins of local minima ({\sc is}), and not by the exploration of
the ``critical'' saddle. As was shown in \cite{sds98,ssdg00},
there indeed exists a separation of time scales between vibrational motion
within an {\sc is}, and hopping between {\sc is}, as envisioned long ago
by Goldstein \cite{g69}. Second, the vanishing of saddles below $T_c$ and
the inexistence of minima above $T_c$ are no longer sharp statements. One
of the primary goal of the present paper is to actually show that the long
time dynamical properties are indeed governed by these rare, deep traps,
{\it even above $T_c$}. The landscape studies of Wales {\em et al.}
\cite{wdmmw00,mw01} and Heuer {\em et al.} \cite{bh00,bh99,dhcm} suggest
that an activated dynamics of a more complex variety, namely escape from
a meta-basin may be taking place already above $T_{c}$.  In this context,
it is interesting to note that some theories of slow dynamics posit that
activated processes are directly responsible for the onset of non-Arrhenius
relaxation at temperatures far above $T_{c}$ \cite{tkv00}.

\begin{figure}
\centering
\includegraphics[width=8cm]{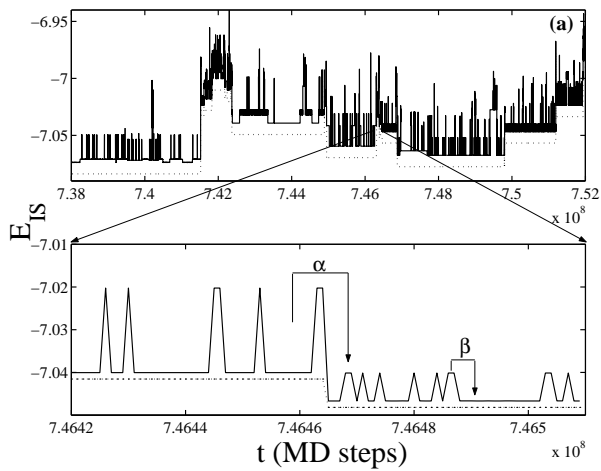}
\caption{{\sc is} (solid line) and meta-basins (dotted
line) visited during the course of a 70 particle MD run at
$T^{*}=0.49$.  The dotted line is slightly shifted along the energy
axis for better clarity. {\sc is} transitions within a
meta-basin are denoted $\beta$ while transitions from one meta-basin
to another are denoted $\alpha$.  The {\sc is} energies are
divided by the total number of particles in the system.}
\end{figure}

A simple way of rationalizing glassy dynamics in the regime where activation
between attractors is the dominant mechanism is provided by the ``trap''
model advocated in \cite{mb96}. In this picture, the dynamics of a coherent
subregion of the system is summarized by the motion of a single point
evolving in a landscape of ``valleys'' or ``traps'', separated by barriers
that can only be overcome via activation.  The wandering of the phase
point is described by a simple master equation with hopping rates that
encode the statistics of barrier heights and the geometry of phase
space.  The simplest of the trap models assumes that the top of the
landscape is flat, and thus the rate of escape from a trap is related
only to the trap {\em depth}. If one further assumes that the probability 
to reach any new trap is identical, the specification of the distribution
of trap energies determines the behavior of {\em all} dynamical observables.
Interestingly,  Ben Arous {\em et al.} \cite{bbg02} have shown that for a 
{\em finite size} p-spin model, the short time dynamics are identical to
that predicted by {\sc mct}, but the long time dynamics are precisely
given by the trap model predictions.  The purpose of this work is to
make concrete connections between the landscape properties of a model
glass forming liquid, and the trap picture above the glass temperature. 
The relevance of the trap picture to describe aging dynamics in the glass 
phase will be discussed in a later work.

The system that we study is the modified 80-20 Lennard-Jones system
studied by Sastry {\em et al.} \cite{sds98}, and we refer the reader to
this work for the details. Since the macroscopic liquid should be viewed
as a partitioning of noninteracting `coherent' cells within which the
landscape picture is sensible \cite{bh00,bh99,dhcm,z83}, we
study small systems containing between 70 and 150 particles. Long (up to
$2 \times 10^{9}$ total time steps) series of inherent structures are
produced from long molecular dynamics runs at various temperatures between
$T^{*}$ = 0.85 and $T^{*}$ = 0.49 ($T_{c} \approx 0.45$).  This corresponds
to a total time run of up to 13 $\mu s$. Quenches are generated at intervals
much shorter than the $\alpha$-relaxation time at a given temperature from
a combined steepest descent and conjugate gradient algorithm. This procedure
is sufficient to resolve nearly all significant inherent structure
transitions.

In Fig.1 we show a portion of an {\sc is} history verses time
($E_{IS}(t)$ vs $t$) for $T^{*}$ = 0.490.  One notes several remarkable
features in this time series \cite{dhcm}.  First, long regions
where the system switches between only a few elementary {\sc is} exist.
Such long-lived collections of {\sc is} clearly provide deep trap states.
Using the meta-basin definition of B\"{u}chner and Heuer \cite{bh00},
the time series of {\sc is} may be mapped into a time series of the visited
meta-basins.  Transitions between {\sc is} within a meta-basin involve very
small flexing of a cage, while transitions between meta-basins (signifying
release from a trap) involve collective rearrangements of particles.
This observation is fully consistent with
Stillinger's landscape picture of the $\beta$ and $\alpha$ processes,
whereby the $\beta$ process corresponds to transitions between {\sc
is} within the meta-basin, while the $\alpha$ process corresponds to
transitions between meta-basins \cite{s95}. Visual inspection
demonstrates clearly that deeper trap states are longer lived, in
accordance with the trap model \cite{mb96}. It is important to note that
although the {\sc is} {\em energy} is used a label of the
instantaneous state of the system, no assumptions are made concerning
the transition dynamics between meta-basins.  In particular,
nontrivial entropic factors may contribute to dynamics of meta-basin
transitions.

\begin{figure}
\centering
\includegraphics[width=8cm]{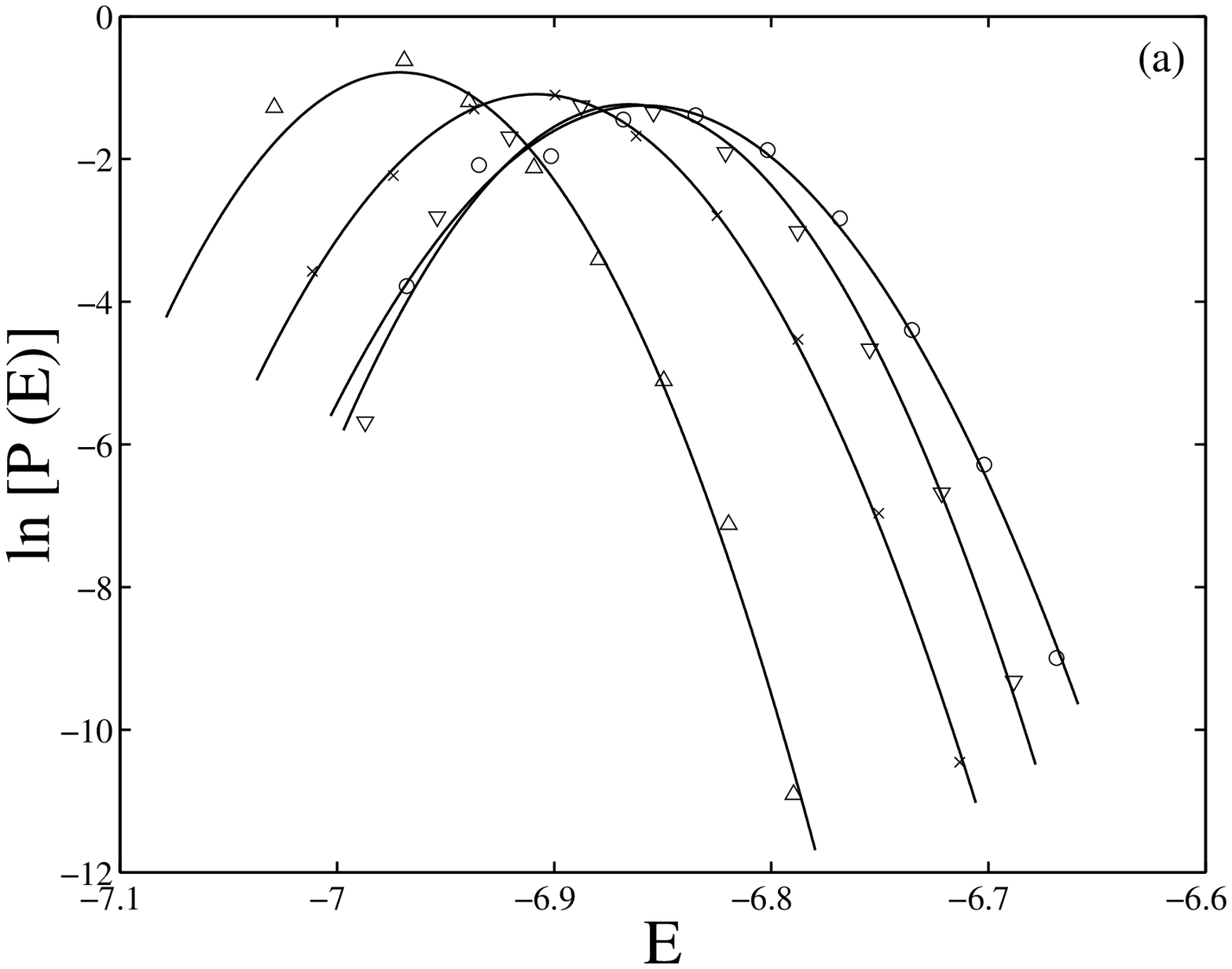}\par
\includegraphics[width=8cm]{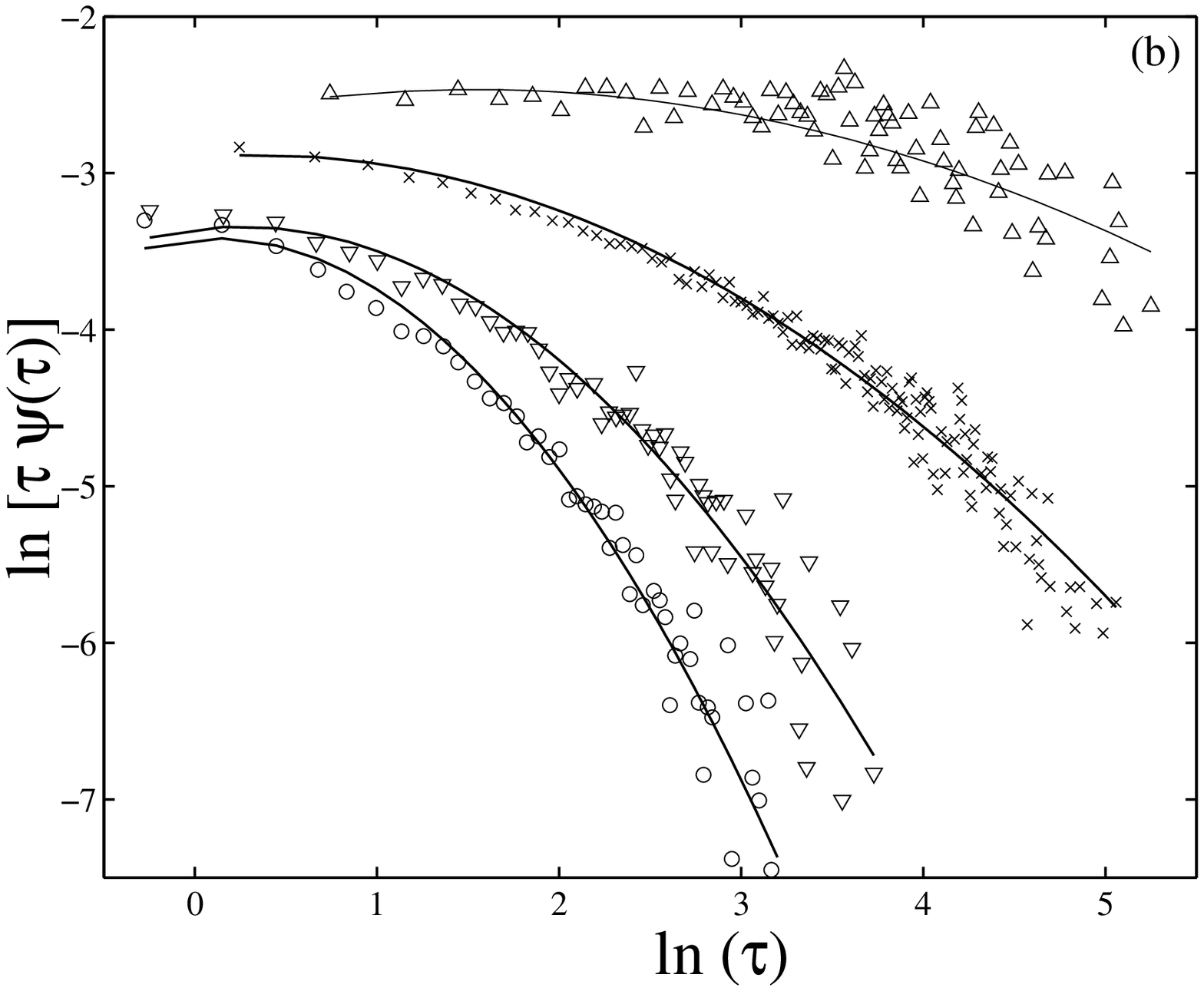}
\caption{(a) Distribution of meta-basin energies (per particle) for
$T^*$ = 0.490, 0.575, 0.669 and 0.764 (from left to right). The solid
lines are the Gaussian fits for the corresponding temperature.
(b) Distribution of hopping times from meta-basins in a 70 particle
system for $T^*$ = 0.764, 0.669, 0.575, and 0.490 (from left to
right). The curves are shifted in time to fit inside the figure.}
\end{figure}

In simple trap models, the fundamental quantity is the distribution of
trap depths.  The most commonly used model assumes an exponential
distribution of trap energies, i.e. $\rho(E)=\left(1/T_{0}\right)\exp
\left(-E/T_{0}\right)$. This model yields a power law distribution of
trapping times $\psi(\tau) \sim \tau^{-(1+\frac{T}{T_{0}})}$, and
a power law correlations for {\em simple} dynamical variables
\cite{mb96,oy00,o95,bs88,ms83}. This model also has a strict glass transition
at the temperature $T_{0}$. Other trap models, such as the Gaussian
model with $\rho(E)=\left(\frac{\exp(-(E-\overline E)^{2}/E_{0}^{2})}
{\sqrt{\pi} E_{0}}\right)$ also display glassy phenomenology, such as
a super-Arrhenius growth of the relaxation time, stretched exponential
decay and (interrupted) aging \cite{mb96}.

Assuming that the meta-basins form traps, the computed distribution of
trap energies may be calculated directly. At all temperatures studied,
the distribution is well fitted by a Gaussian, as shown in Fig.2(a).
Previous studies have demonstrated that the distribution of inherent
structures ({\em not} meta-basins) is Gaussian \cite{bh99,s01,skt99}.

\begin{figure}
\centering
\includegraphics[width=8cm]{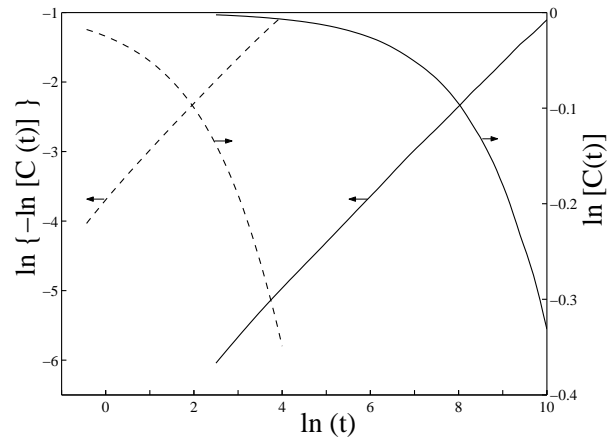}
\caption{Correlation function $C(t)$ as defined in the text. Both
power law and stretched exponential fits are shown.  The data clearly
supports the stretched exponential fit over the power law fit. The
solid line is for $T^*$ = 0.490 and dotted line is for $T^*$=0.575.
}
\end{figure}

Given simple activated dynamics to leave a trap of depth $E$, i.e.
trap lifetimes $\tau = \tau_{0}\exp(\beta E)$, the distribution of
trapping times may be computed.  For the Gaussian trap model, this
yields a distribution of trapping times $\psi(\tau)=\frac{1}{\tau
\sqrt{\pi}\Delta}\exp\left(-\left(\frac{\ln(\tau/\tau_{0})}{\Delta}
\right)^{2}\right)$, where $\Delta \equiv \beta E_{0}$. Note that the
product $\tau \psi(\tau)$ is log-normal. In Fig.2(b) we show fits of
$\ln(\tau \psi(\tau))$ vs. $\ln(\tau)$. The log-normal behavior predicted
by the Gaussian trap model fits remarkably well over a wide range of time
scales.  The attempt time $\tau_{0}$ is determined to be on the order
of $\sim 1ps$, reasonable for the ``vibrational'' prefactor of an
activated process in a dense Lennard-Jones liquid.  Given that
meta-basins can have lifetimes spanning up to a fraction of one $\mu
s$, there clearly exists a separation of time scales between inter-
and intra-trap (meta-basin) motion.  This timescale separation does
not exist at the level of single {\sc is} above $T_{c}$.
It is interesting to note that previous computational studies have
determined {\em power law} waiting time distributions in a variety of
systems \cite{dhcm,adg99,hh99}.  On the basis of a plot of $\ln\psi(\tau)$
vs. $\ln(\tau)$ it is difficult to detect the distinction between the
power law and log-normal versions of $\psi(\tau)$, while plots of
$\ln(\tau \psi(\tau))$ vs. $\ln(\tau)$ clearly reveal this distinction.
The Gaussian widths $\Delta$ determined from fits of $\psi(\tau)$ are
larger than $\beta E_{0}$ determined from $\rho(E)$ directly. This fact
shows that one of the crude assumptions of the trap model cannot be taken
literally; for example the energy depth of the meta-basin alone does not
determine the absolute barrier for trap escape.  Interestingly, the
nontrivial entropic and saddle point dependence of $E_{0}$ only
quantitatively, and not qualitatively affects estimates of the
distribution of trapping times.

We now compare simple dynamical predictions of the Gaussian trap model
to computer simulations in the Lennard-Jones mixture. The basic quantity
investigated is $C(t)=\langle \delta \epsilon_{T}(t)\delta \epsilon_{T}
(0)\rangle$, namely, the correlation function of fluctuations of visited
meta-basin depths.  In the exponential trap model, $C(t)$ decays as a
power law, while for the Gaussian model the behavior of $C(t)$ may be
{\em approximated} with a stretched exponential decay \cite{mb96}: $C(\tau)
\sim \exp\left(-(a(T)t^{b(T)})\right)$, with a stretching coefficient
$b(T) \sim \left(1+\frac{1}{2}(\Delta)^{2}\right)^{-\frac{1}{2}}$. For
the range of temperatures studied, the Gaussian trap prediction of
stretched exponential relaxation is reasonably well borne out, as is
shown in Fig.3.  Approximate stretched exponential dependence is seen
for all temperatures $T^{*} \leq 1$.  At the two lowest temperatures
studied in this work ($T^{*}$ = 0.575 and $T^{*}$ = 0.490) where the
extracted values of $b(T)$ are most accurate, the extracted values of
$b(T)$ are $b(0.575)$ = 0.68$\pm$0.1 and $b(0.490)$ = 0.61$\pm$0.05,
respectively.  The values predicted from the trap model are $b(0.575)$
= 0.45$\pm$0.05 and $b(0.490)$ = 0.41$\pm$0.1, respectively. Thus, the
predicted values are in reasonable agreement with the values extracted
from simulation \cite{note}.

Our results show that activated process are important even above $T_c$
in supercooled liquids. This does does not contradict the applicability
of {\sc mct} to describe short time dynamics in this regime, nor does it
contradict the results of \cite{adrss00,bbczg00,c01}, which show that
{\em most} accessible saddles are unstable above $T_c$. The long time
dynamics, however, will be dominated by rare, deep traps. Our work is
consistent with this, provided that traps (meta-basins) smoothly
transition from a collection of connected inherent structures above
$T_{c}$ to {\em single} inherent structures below $T_{c}$.  That is,
above $T_{c}$ the barriers separating the inherent structures inside
a meta-basin are small compared to temperature, while below $T_{c}$
they are large. Computer simulation evidence supports this picture
\cite{sds98}. This would imply that the simple ``single-level'' trap
model above $T_{c}$ should become a ``multi-level'' trap picture below
$T_{c}$ \cite{spin98}, where interesting aging effects should then take
place as in spin-glasses \cite{rmb01}.

We gratefully acknowledge useful discussions with E. Bertin,
B. Chakraborty, D. Das, I. Giardina, C. Godr\`eche, A. Heuer,
J. Klafter, J. Kondev, and R. Yamamoto.  D.R.R. and R.A. Denny
acknowledge the NSF for financial support.


\begin{thebibliography}{15truecm}
\bibitem{note} While the simulated values of $b(T)$ and those predicted
based on the Gaussian trap model with $\Delta$ extracted from
$\psi(\tau)$ are in reasonable agreement (within the error bars),
there is systematic tendency for the trap prediction to underestimate
the value of $b(T)$.  For large values of $\tau$ our statistics for
$\psi(\tau)$ are poor, however there does seem to be a slight systematic
deviation from log-normal behavior for the largest $\tau$'s (not shown
in Fig. 2(b)).  Including this curvature change results in slightly
larger estimates of $b(T)$. Doliwa and Heuer (private discussion) have
demonstrated that their data is consistent with a Gaussian trap model,
but with saddle-point values that depend somewhat on the trap depth.
This observation is sufficent to explain the small deviations
mentioned above.  See also the related discussion in \cite{xw01}.
\end{thebibliography}
\end{document}